\documentclass{article}%
\usepackage{amsmath}
\usepackage{amsfonts}
\usepackage{amssymb}
\usepackage{graphicx}%
\begin{document}

\title{Isotropy, shear, symmetry and exact solutions for relativistic fluid spheres}
\author{Ron Wiltshire\\The Division of Mathematics \& Statistics,\\The University of Glamorgan,\\Pontypridd CF37 1DL, UK\\email: rjwiltsh@glam.ac.uk}
\maketitle

\begin{abstract}
The symmetry method is used to derive solutions of Einstein's equations for
fluid spheres using an isotropic metric and a velocity four vector that is
non-comoving. Initially the Lie, classical approach is used to review and
provide \ a connecting framework for many comoving and so shear free
solutions. This provides the basis for the derivation of the classical point
symmetries for the more general and mathematicaly \ less tractable description
of Einstein's equations in the non-comoving frame. Although the range of
symmetries is restrictive, existing and new symmetry solutions with non-zero
shear are derived. The range is then extended using the non-classical direct
symmetry approach of Clarkson and Kruskal and so additional new solutions with
non-zero shear are also presented. The kinematics and pressure, energy
density, mass function of these solutions are determined.

\medskip PACs numbers: 0420, 0440, 0230, 0270

\end{abstract}

\section{Introduction}

Solutions of Einstein's field equations that describe the motion of
spherically symmetric fluid spheres have been discussed by many authors for
many reasons which include their wide ranging applications to cosmology and
astrophysics. The most general solutions may be obtained using a metric in
isotropic form for which the four velocity is non-comoving and which possess
non-zero shear. \ However, such cases are in general far less mathematically
tractable than their more restrictive comoving counterparts. One of the most
important techniques available in the derivation of comoving, or shear free,
solutions is the Lie, classical point symmetry approach \ although the method
has not been applied to non-comoving descriptions and non-classical or non-Lie
approaches are seldom applied to the problem of solving Einstein's equations.

The Lie approach applied to comoving systems was originally described by
Kustaanheimo and Qvist \cite{ku} and by Stephani \cite{sti}, Stephani and Wolf
\cite{stw} \ who present the full range of symmetries and show that these Lie
solutions of Einstein's equations contain,two arbitrary functions of time. Of
course there are many other known solutions involving only one arbitrary
function of time perhaps most notably by McVittie \cite{MCV33,mc}, but also by
for example Nariai \cite{nr} and Chakravarty \cite{ch}. Whilst the
mathematical links between the many solutions have been comprehensively
discussed by for example McVittie \cite{mc}, Srivastava \cite{sr}, Sussman
\cite{sus} \ and Stephani \textit{et al} \cite{stepbook} \ it is interesting
to note that these have not been discussed within the overall framework of the
symmetry method. Ultimately, of course it is the physical properties of these
solutions of considerable interest

Solution's of Einstein's equations for fluid spheres described by a non-
comoving observer on the other hand are still relatively unknown inspite of
recent interest by for example Ishak and Lake \cite{la}, Ishak \cite{is},
Senovila and Vera \cite{se} and Davision \cite{da}. Although such solutions
often have kinematics rich in shear, acceleration and expansion the
mathematical expression of the condition for isotropy of pressure is
mathematically far less tractable than for a comoving system. Indeed the
solutions of Narlikar \& Moghe \cite{na}, McVittie and Wiltshire \cite{wt}
discussed by Knutsen \cite{kn1,br,kn} were found with the aid of ad hoc
mathematical assumptions. Clearly the symmetry approach \ provides a more
structured environment in which to derive and discuss solutions of Einstein's
equations with a non-comoving four velocity.

It will be shown however that the complexity of Einstein's equations in the
non-comoving frame reduces the range of Lie symmetries available compared with
the comoving case. It follows that non-Lie approaches to symmetry such as the
non-classical method of Bluman and Cole \cite{blcl} and the direct method of
Clarkson and Kruskal \cite{clk} \ should also be investigated to extend the
range of similarity solutions.. It will be seen that the direct method based
upon symmetries for comoving systems results in new solutions.

Whist the focus here concerns the symmetry framework for solutions of
Einstein's equations it is important to stress that the physical properties of
all solutions must be established. For comoving cases these results are well
known, see for example, McVittie \cite{mc1966}, Knutsen \cite{kn1986}, Sussman
\cite{sus,s1988a,s1988b} whilst Knutsen \cite{kn1,kn} has examined some
non-comoving solutions. A review of many physical properties has been
discussed by Krasinski \cite{kras}.

This paper is organised in the following way. Einstein's equations, in terms
of the isotropy condition for fluid spheres, are presented for the
non-comoving and comoving four velocity in section 2. The kinematic parameters
are evaluated in general terms. Section 3 describes the Lie symmetry approach,
used in section 4 to review symmetry solutions for comoving cases.The
interrelationship between known solutions of Einstein's equations is also
discussed. The Lie symmetry approach is extended to non-comoving cases in
section 5 with a new solution presented. In section 6 non-Lie, non-comoving
solutions with non-zero shear are derived using the direct approach of
Clarkson and Kruskal and further new solutions, with their kinematic
properties is given.

\section{Fluid spheres, isotropy, kinematic parameters}

\subsection{Preliminary equations}

It is the intention here to consider the isotropic coordinate system for
which\qquad%
\begin{equation}
d\sigma^{2}=e^{2\lambda}dt^{2}-e^{2\mu}\left(  dr^{2}+r^{2}d\Omega^{2}\right)
\qquad\qquad d\Omega^{2}=d\theta^{2}+\sin^{2}\left(  \theta\right)  d\phi^{2}
\label{raw6}%
\end{equation}
where $\lambda=\lambda\left(  r,t\right)  $, $\mu=\mu\left(  r,t\right)  $. In
addition Einstein's field equations will be taken to be%
\begin{equation}
G_{k}^{i}=-8\pi T_{k}^{i}\qquad\qquad T_{k}^{i}=\left(  \rho+p\right)
u^{i}u_{k}-\delta_{k}^{i}p \label{pl59}%
\end{equation}
where $G_{k}^{i}$ is \ the Einstein tensor and $T_{k}^{i}$ is the energy
momentum tensor of the fluid sphere. The pressure $p$, energy density $\rho$
and mass function $m$,see for example, Misner and Sharp \cite{mi}, Cahill and
McVittie \cite{ca} may be calculated using:%

\begin{equation}
8\pi p=G_{2}^{2}\qquad8\pi\rho=G_{2}^{2}-G_{1}^{1}-G_{4}^{4}\qquad
m=\frac{re^{\mu}}{2}\left\{  1+e^{2\left(  \mu-\lambda\right)  }r^{2}\mu
_{t}^{2}-\left(  1+r\mu_{r}\right)  ^{2}\right\}  \label{is40}%
\end{equation}
where the suffix $r$, $t$ indicates a partial derivative. $\ $

Furthermore $u^{i}$ are the components of the velocity four vector that
satisfy $u^{i}u_{i}=1$ and where $u^{2}=0=u^{3}$. These may be used to
calculate the kinematic characteristics of the fluid \cite{stepbook}. In
particular the expansion $\Theta$, acceleration $\mathring{u}_{i},$ shear
tensor $\sigma_{ik}$ and shear invariant $\sigma$ are respectively:%
\begin{align}
\Theta &  =u_{;i}^{i}\qquad\qquad\mathring{u}_{i}=u_{i;j}u^{j}\nonumber\\
\sigma_{ik}  &  =u_{\left(  i;k\right)  }-\mathring{u}_{(i}u_{k)}-\frac{1}%
{3}h_{ik}\Theta\qquad\qquad\sigma=\sigma_{ik}\sigma^{ik} \label{xz12}%
\end{align}
where $h_{ik}=g_{ik}-u_{i}u_{k}$ is the projection tensor and the parentheses
indicate symmetrisation so that $u_{\left(  i;k\right)  }=\frac{1}{2}\left(
u_{i;k}+u_{k;i}\right)  $.

Using Walker \cite{walker} the assumption of spherical symmetry for a perfect
fluid leads to the isotropy condition:%
\begin{equation}
\Delta=e^{2\lambda}\left(  G_{1}^{4}\right)  ^{2}+e^{2\mu}\left(  G_{2}%
^{2}-G_{1}^{1}\right)  \left(  G_{2}^{2}-G_{4}^{4}\right)  =0 \label{is5}%
\end{equation}

\subsection{Non-comoving (non-zero shear) cases}

To determine the isotropy condition and kinematic parameters for noncomoving
cases $u^{1}\neq0$ it is convenient to write:%
\begin{align}
G_{2}^{2}-G_{1}^{1}  &  =e^{-2\mu}K_{1}\qquad\qquad\qquad e^{2\lambda}%
G_{1}^{4}=-e^{2\mu}G_{4}^{1}=2K_{3}\nonumber\\
G_{2}^{2}-G_{4}^{4}  &  =-2e^{-2\lambda}K_{4}-e^{-2\mu}K_{2} \label{id10a}%
\end{align}
In this way the isotropy condition (\ref{is5}) may be written as:%
\begin{equation}
\Delta=e^{2\left(  \mu-\lambda\right)  }\left\{  4K_{3}^{2}-2K_{1}%
K_{4}\right\}  -K_{1}K_{2}=0 \label{id35}%
\end{equation}
where $K_{1}$,$K_{2}$,$K_{3}$ and $K_{4}$ are respectively:%
\begin{equation}%
\begin{tabular}
[c]{ll}%
$K_{1}=\mu_{rr}+\lambda_{rr}+\lambda_{r}^{2}-\mu_{r}^{2}-2\lambda_{r}\mu
_{r}-\dfrac{\left(  \mu_{r}+\lambda_{r}\right)  }{r}$ & $\qquad K_{3}=\mu
_{rt}-\lambda_{r}\mu_{t}$\\
$K_{2}=\mu_{rr}-\lambda_{rr}-\lambda_{r}^{2}+\mu_{r}^{2}+\dfrac{\left(
3\mu_{r}-\lambda_{r}\right)  }{r}$ & $\qquad K_{4}=\mu_{tt}-\lambda_{t}\mu
_{t}$%
\end{tabular}
\ \ \ \ \label{id20b}%
\end{equation}
From (\ref{id10a}) and (\ref{pl59}) and taking $u^{4}$ to be positive it may
be shown that%
\begin{equation}
u^{1}\equiv f=-\,\frac{K_{1}g}{2K_{3}}\,e^{2\,\lambda-2\,\mu}\qquad
u^{4}\equiv g=\frac{e^{-\lambda}}{\sqrt{1-\left(  \frac{K_{1}}{2K_{3}}\right)
^{2}e^{2\left(  \lambda-\mu\right)  }}} \label{xz50}%
\end{equation}
Thus after some calculation the expansion is%
\begin{equation}
\Theta=f\lambda_{r}+g\lambda_{t}+\frac{2f\left(  \mu_{r}r+1\right)  }{r}%
+f\mu_{r}+3g\mu_{t}+g_{t}+f_{r} \label{xz63}%
\end{equation}
whilst the two non-zero components of acceleration are%
\begin{equation}
\mathring{u}_{1}=-\frac{g\mathring{u}_{4}}{f}=-e^{2\mu}\left(  f^{2}%
\lambda_{r}+f^{2}\mu_{r}+2fg\mu_{t}+f_{t}g+ff_{r}\right)  -\lambda_{r}
\label{xz66}%
\end{equation}
The non-zero components of the shear tensor are:%
\begin{equation}
\sigma_{11}=-\frac{2g^{2}e^{2\,\lambda+2\,\mu}\left(  fr\lambda_{r}%
+f_{r}r-f\right)  }{3r}-\frac{2fge^{4\mu}\left(  f\mu_{t}+f_{t}\right)  }{3}
\label{xz72}%
\end{equation}%
\begin{equation}
\sigma_{22}=\frac{\sigma_{33}}{\sin^{2}\theta}=-\frac{\sigma_{11}%
\,r^{2}\,e^{-{2\,\lambda}}}{2\,g^{2}} \label{xz72d}%
\end{equation}%
\begin{equation}
\sigma_{44}=\frac{\,\sigma_{11}e^{-{2\,\mu}}\,\left(  g^{2}\,e^{2\lambda
}-1\right)  }{g^{2}}\qquad\qquad\sigma_{14}=\sigma_{41}=-\frac{\sigma_{11}%
\,f}{g} \label{xz82}%
\end{equation}
and finally the shear invariant is%
\begin{equation}
\sigma=\frac{3\sigma_{11}^{2}e^{-4\left(  \,\lambda+\,\mu\right)  }}{2\,g^{4}}
\label{xz90}%
\end{equation}

Equations (\ref{xz72}) to (\ref{xz90}) demonstrate that each component of the
shear tensor is proportional to $\sigma_{11}$whilst the shear invariant is
proportional to $\sigma_{11}^{2}$. Hence it is necessary only to calculate
$\sigma_{11}$ \ to determine whether solutions of Einstein's equations have
non-zero shear.

\subsection{Comoving (shear-free) cases}

In the case of a comoving observer $u^{1}\equiv f=0$ and$\ u^{4}\equiv
g=e^{-\lambda}$ and so from (\ref{xz63}) to (\ref{xz90}) the only non-zero
kinematic parameters are:%
\begin{equation}
\Theta=3\mu_{t}e^{-\lambda}\qquad\mathring{u}_{1}=-\lambda_{r} \label{xz23}%
\end{equation}
and so the shear tensor is necessarily zero for metrics isotropic in form. For
comoving systems the condition for isotropy of pressure is the particular
solution of (\ref{id35}) that gives rise to the two equations:%
\begin{equation}
\Delta_{1}=K_{3}=0\qquad\Delta_{2}=K_{1}=0 \label{ba6}%
\end{equation}

\section{Classical symmetry method}

Initially it is the intention to discuss solutions of these equations in the
context of Lie's classical symmetry. In this method one-parameter
infinitesimal point transformations,with group parameter $\varepsilon$ are
applied to the dependent and independent variables $\left(  r,t,\lambda
,\mu\right)  $. In this case the transformations are%
\begin{align}
\bar{r}  &  =r+\varepsilon\eta_{1}\left(  r,t,\lambda,\mu\right)  +O\left(
\varepsilon^{2}\right)  \qquad\bar{t}=t+\varepsilon\eta_{2}\left(
r,t,\lambda,\mu\right)  +O\left(  \varepsilon^{2}\right) \label{trans1}\\
\bar{\lambda}  &  =\lambda+\varepsilon\phi_{1}\left(  r,t,\lambda,\mu\right)
+O\left(  \varepsilon^{2}\right)  \qquad\bar{\mu}=\mu+\varepsilon\phi
_{2}\left(  r,t,\lambda,\mu\right)  +O\left(  \varepsilon^{2}\right) \nonumber
\end{align}
and the Lie method, see for example Stephani, \cite{st} or Bluman and Kumei,
\cite{kum}, requires form invariance of the solution set:%
\begin{equation}
\Sigma\equiv\left\{  \lambda\left(  r,t\right)  ,\mu=\mu\left(  r,t\right)
,\text{ }\boldsymbol{\Delta=0}\right\}  \label{cond1}%
\end{equation}
where $\boldsymbol{\Delta=}\left\{  \Delta_{1},\Delta_{2}\right\}  $ is
defined by (\ref{ba6}) for a comoving observer and $\boldsymbol{\Delta
=}\left\{  \Delta\right\}  $ defined by (\ref{id35}) in the non-comoving case.
This results in a system of overdetermined, linear equations for the
infinitesimals $\eta_{1}$, $\eta_{2}$, $\phi_{1}$ and $\phi_{2}$.. The
corresponding Lie algebra of symmetries is the set of vector fields%
\begin{equation}
\mathcal{X}=\eta_{1}\left(  r,t,\lambda,\mu\right)  \frac{\partial}{\partial
r}+\eta_{2}\left(  r,t,\lambda,\mu\right)  \frac{\partial}{\partial t}%
+\phi_{1}\left(  x,t,\lambda,\mu\right)  \frac{\partial}{\partial\lambda}%
+\phi_{2}\left(  r,t,\lambda,\mu\right)  \frac{\partial}{\partial\mu}
\label{gen1}%
\end{equation}
The condition for invariance of (\ref{ba6}) and (\ref{id35}) is the equation%
\begin{equation}
\mathcal{X}_{E}^{(2)}\left(  \boldsymbol{\Delta}\right)  \mathbf{\mid
}_{\boldsymbol{\Delta=0}}=0 \label{inv1}%
\end{equation}
where the second prolongation operator $\mathcal{X}_{E}^{(2)}$ is written in
the form%
\begin{align}
\mathcal{X}_{E}^{(2)}  &  =\mathcal{X}+\phi_{1}^{\left[  t\right]  }%
\frac{\partial}{\partial\lambda_{t}}+\phi_{1}^{\left[  r\right]  }%
\frac{\partial}{\partial\lambda_{r}}+\phi_{1}^{\left[  tt\right]  }%
\frac{\partial}{\partial\lambda_{tt}}+\phi_{1}^{\left[  rt\right]  }%
\frac{\partial}{\partial\lambda_{rt}}+\phi_{1}^{\left[  rr\right]  }%
\frac{\partial}{\partial\lambda_{rr}}\nonumber\\
&  +\phi_{2}^{\left[  t\right]  }\frac{\partial}{\partial\mu_{t}}+\phi
_{2}^{\left[  r\right]  }\frac{\partial}{\partial\mu_{r}}+\phi_{2}^{\left[
tt\right]  }\frac{\partial}{\partial\mu_{tt}}+\phi_{2}^{\left[  rt\right]
}\frac{\partial}{\partial\mu_{rt}}+\phi_{2}^{\left[  rr\right]  }%
\frac{\partial}{\partial\mu_{rr}} \label{prola}%
\end{align}
where $\phi_{i}^{[t]}$, $\phi_{i}^{[r]}$,$\phi_{i}^{[rr]}$,$\phi_{i}^{[rt]}$
and $\phi_{i}^{[tt]}$ , $i=1,2$ are defined through the transformations of the
first and second partial derivatives of $\lambda$ and $\mu$ to the first order
in $\varepsilon$. Thus for example:%
\begin{align}
\bar{\lambda}_{\bar{r}}  &  =\lambda_{r}+\varepsilon\phi_{1}^{[r]}\left(
r,t,\lambda,\mu\right)  \qquad\quad\bar{\mu}_{\bar{r}}=\mu_{r}+\varepsilon
\phi_{2}^{[r]}\left(  r,t,\lambda,\mu\right) \nonumber\\
\bar{\lambda}_{\bar{t}}  &  =\lambda_{t}+\varepsilon\phi_{1}^{[t]}\left(
r,t,\lambda,\mu\right)  \qquad\quad\bar{\mu}_{\bar{t}}=\mu_{t}+\varepsilon
\phi_{2}^{[t]}\left(  r,t,\lambda,\mu\right)  \label{mhq}%
\end{align}
Once the infinitesimals are determined the symmetry variables may be found
from condition for surface invariance applied to $\lambda=\lambda\left(
r,t\right)  $, $\mu=\mu\left(  r,t\right)  $:%
\begin{equation}
\Omega_{1}\mathbf{=}\phi_{1}\mathbf{-}\eta_{1}\lambda_{r}-\eta_{2}\lambda
_{t}=0\qquad\qquad\Omega_{2}\mathbf{=}\phi_{2}\mathbf{-}\eta_{1}\mu_{r}%
-\eta_{2}\mu_{t}=0 \label{surf2}%
\end{equation}
Both \texttt{Maple }and the now unsupported \texttt{Macsyma} software have
been used to calculate the determining equations.

\section{Review of comoving (shear-free) solutions}

\subsection{Symmetry reduction}

The determining equations resulting from the invariance condition (\ref{inv1})
applied to (\ref{ba6}) give the following solutions for the infinitessimals
$\eta_{1}$,$\eta_{2}$, $\phi_{1}$ and $\phi_{2}$ in the generator
(\ref{gen1}):
\begin{equation}%
\begin{tabular}
[c]{ll}%
$\eta_{1}=\eta_{1}\left(  r\right)  =\dfrac{c_{3}}{r}+c_{2}r-\dfrac{c_{1}%
r^{3}}{2}$ & $\qquad\phi_{1}=\phi_{1}\left(  t\right)  $\\
$\eta_{2}=\eta_{2}\left(  t\right)  \equiv\eta\left(  t\right)  $ &
$\qquad\phi_{2}=c_{0}+c_{1}r^{2}$%
\end{tabular}
\ \ \ \ \label{arb6a}%
\end{equation}
where $c_{0}$, $c_{1}$, $c_{2}$ and $c_{3}$ are constant symmetry parameters.
Integrating the surface conditions (\ref{surf2}) gives%

\begin{equation}
\lambda=\Psi\left(  \omega\right)  +\ln A\left(  t\right)  \qquad\qquad
\mu=\Phi\left(  \omega\right)  +\ln\left(  h\left(  r\right)  \right)
\label{arb23}%
\end{equation}%
\begin{equation}
\omega=\int\frac{dr}{\frac{c_{3}}{r}+c_{2}r-\frac{c_{1}r^{3}}{2}}-\int
\frac{dt}{\eta\left(  t\right)  }\qquad\ln\left(  h\left(  r\right)  \right)
=\int\frac{c_{0}-c_{2}+c_{1}r^{2}}{\frac{c_{3}}{r}+c_{2}r-\frac{c_{1}r^{3}}%
{2}}dr \label{arb24}%
\end{equation}
and $\ A\left(  t\right)  $ is arbitrary. Using (\ref{arb23}) , (\ref{arb24})
and the functions $z=z\left(  \omega\right)  $, $Z=Z\left(  \omega\right)  $
equation (\ref{ba6}) is satisfied by the two ordinary differential equations
\begin{equation}
z\equiv e^{\Psi}=\Phi_{\omega}\equiv-\frac{Z_{\omega}}{Z} \label{barb15}%
\end{equation}%
\begin{equation}
z_{\omega\omega}-zz_{\omega}-2c_{0}z_{\omega}-z^{3}-2c_{0}z^{2}+\left(
2c_{1}c_{3}+c_{2}^{2}-c_{0}^{2}\right)  z=0 \label{barb30}%
\end{equation}
and is the equation used by McVittie \cite{mc}. Thus his solutions are
classical Lie solutions of the coupled system (\ref{ba6}). However
(\ref{barb30}) is also equivalent to:%
\begin{equation}
Z_{\omega\omega}-2c_{0}Z_{\omega}+bZ^{2}-\left(  2c_{1}c_{3}+c_{2}^{2}%
-c_{0}^{2}\right)  Z=0 \label{kust12x}%
\end{equation}
where $b$ is a constant. This equation may be generalised by referring to
Kustaanheimo and Qvist \cite{ku} who showed that solutions of Einstein's
equations may be expressed in terms of $L=L\left(  x,t\right)  $:%
\begin{equation}
L=e^{-\mu}\qquad x=r^{2}\qquad e^{\lambda}=A\left(  t\right)  \mu
_{t}=-A\left(  t\right)  \frac{L_{t}}{L} \label{kus49}%
\end{equation}
so that (\ref{ba6}) reduces to
\begin{equation}
L_{xx}=L^{2}F\left(  x\right)  \label{kus51}%
\end{equation}
Substituting the second of (\ref{arb23}) into (\ref{kus51}) it follows that:%
\begin{equation}
F\left(  x\right)  =-b\frac{h\left(  x\right)  }{\left(  c_{1}x^{2}%
-2c_{2}x-2c_{3}\right)  ^{2}}=-b\frac{e^{-c_{0}\int\frac{dx}{c_{1}x^{2}%
-2c_{2}x-2c_{3}}}}{\left(  c_{1}x^{2}-2c_{2}x-2c_{3}\right)  ^{\frac{5}{2}}}
\label{kus15}%
\end{equation}
and (\ref{kust12x}) is again recovered with $Z$ defined by (\ref{barb15}).
Equation (\ref{kus15}) is a general form for $F\left(  x\right)  $ and has
been reported by Stephani and Wolf \cite{stw} in their symmetry analysis of
(\ref{kus51}). Thus the Lie analysis of (\ref{kus51}) admits solutions
containing two arbitrary functions of time whilst solutions obtained by a
symmetry analysis of the coupled system have only one arbitrary function, the
similarity variable. The latter are the McVittie metrics which are otherwise
identical. Hence $Z=Z\left(  \omega\right)  $ in (\ref{kust12x}) may be
replaced by $\tilde{Z}=$ $\tilde{Z}\left(  \omega,t\right)  $ which satisfies:%
\begin{equation}
\tilde{Z}_{\omega\omega}-2c_{0}\tilde{Z}_{\omega}+b\tilde{Z}^{2}-\left(
2c_{1}c_{3}+c_{2}^{2}-c_{0}^{2}\right)  \tilde{Z}=0 \label{kus12}%
\end{equation}
so that%
\begin{equation}
e^{\lambda}=-A\left(  t\right)  \frac{\tilde{Z}_{t}}{\tilde{Z}}\qquad e^{\mu
}=L^{-1}\left(  x,t\right)  \equiv h\left(  x\right)  \tilde{Z}^{-1}\left(
\omega,t\right)  \label{vc51}%
\end{equation}

\subsection{Relationships between $\omega\left(  x,t\right)  $ and $h\left(
x\right)  $}

Equation (\ref{arb24}) gives rise to the particular cases that are presented
in table 1

\begin{center}%
\[%
\begin{tabular}
[c]{|c|c|c|c|c|}\hline\hline
\multicolumn{1}{||c|}{} & \multicolumn{1}{||c|}{$h\left(  x\right)  $} &
\multicolumn{1}{||c|}{$\omega\left(  x,t\right)  $} &
\multicolumn{1}{||c|}{$H\left(  x\right)  $} & \multicolumn{1}{||c||}{%
\begin{tabular}
[c]{l}%
\textit{Para-}\\
\multicolumn{1}{c}{\textit{meters }$c_{i}$}%
\end{tabular}
}\\\hline\hline
\textit{1} & $e^{-\dfrac{c_{o}x}{2c_{3}}}$ & $\frac{x}{2c_{3}}-\int\frac
{dt}{\eta\left(  t\right)  }$ &  &
\begin{tabular}
[c]{l}%
$c_{1}=0$\\
$c_{2}=0$%
\end{tabular}
\\\hline
\textit{2} & $H^{\dfrac{c_{0}-c_{2}}{2c_{2}}}$ &
\begin{tabular}
[c]{l}%
$\frac{1}{2c_{2}}\ln\left\{  -H\right\}  $\\
$-\int\frac{dt}{\eta\left(  t\right)  }$%
\end{tabular}
& $2\left(  c_{2}x+c_{3}\right)  $ &
\begin{tabular}
[c]{l}%
$c_{1}=0$\\
$c_{2}\neq0$%
\end{tabular}
\\\hline
\textit{3} & $\dfrac{e^{c_{0}H}}{\sqrt{c_{1}x^{2}-2c_{2}x-2c_{3}}}$ &
$H-\int\frac{dt}{\eta\left(  t\right)  }$ &
\begin{tabular}
[c]{l}%
$\dfrac{\ln\left\{  \frac{c_{1}x+\sqrt{\beta}-c_{2}}{c_{1}x-\sqrt{\beta}%
-c_{2}}\right\}  }{2\sqrt{\beta}}$%
\end{tabular}
&
\begin{tabular}
[c]{c}%
$c_{1}\neq0$\\
$\beta=c_{2}^{2}+$\\
$2c_{1}c_{3}>0$%
\end{tabular}
\\\hline
\textit{4} & $\dfrac{e^{c_{0}H}}{\sqrt{c_{1}x^{2}-2c_{2}x-2c_{3}}}$ &
$H-\int\frac{dt}{\eta\left(  t\right)  }$ & $-\dfrac{\arctan^{\left\{
\frac{c_{1}x-c_{2}}{\sqrt{-\beta}}\right\}  }}{\sqrt{-\beta}}$ &
\begin{tabular}
[c]{c}%
$c_{1}\neq0$\\
$\beta=c_{2}^{2}+$\\
$2c_{1}c_{3}<0$%
\end{tabular}
\\\hline
\textit{5} & $He^{Hc_{0}}$ & $H-\int\frac{dt}{\eta\left(  t\right)  }$ &
$\dfrac{1}{c_{1}x-c_{2}}$ &
\begin{tabular}
[c]{c}%
$c_{1}\neq0$\\
$\beta=c_{2}^{2}+$\\
$2c_{1}c_{3}=0$%
\end{tabular}
\\\hline
\end{tabular}
\]

\textit{Table 1: Relationship between }$h\left(  x\right)  $ and
$\omega\left(  x,t\right)  $ with $x=r^{2}$
\end{center}

Note that when $c_{1}=0$ , $c_{0}=0$ then from table 1 and (\ref{kus15})
$F\left(  x\right)  $ is constant giving the results of Wyman \cite{wy1}.
However when $c_{1}=0$, $c_{0}=-5c_{2}/7$ then $F\left(  x\right)  =-b\left\{
-2\left(  c_{2}x+c_{3}\right)  \right\}  ^{-\frac{20}{7}}$giving the result of
Wyman \cite{wy}. Also when $c_{0}=5c_{2}/7$, $c_{3}=0$, $c_{1}>0$ then from
(\ref{kus15}) $F\left(  x\right)  =-bc_{1}^{\frac{5}{14}}x^{-\frac{15}{7}%
}\left(  c_{1}x-2c_{2}\right)  ^{-\frac{20}{7}\text{ }}$ which gives the
result of Srivastava \cite{sr}.

\subsection{Summary of solutions for $\tilde{Z}\left(  \omega,t\right)  $}

Consider now solutions of (\ref{kus12}) required for the metric (\ref{vc51}).

Firstly when $b=0$ then by (\ref{kus15}) $F\left(  x\right)  =0$ and
(\ref{kus51}) gives the simple form $L\left(  x,t\right)  =A\left(  t\right)
x+B\left(  t\right)  $. Moreover (\ref{kust12x}) or (\ref{barb15}) with $b=0$
\ is wholly equivalent as has been discussed by McVittie \cite{mc} and
Srivastava \cite{sr}.

In the case when $b\neq0$ the solutions of (\ref{kus51}) are summarised in
Table 2. This table expresses several solutions in terms of Weierstrass
$P$-functions, $P\left(  y,t\right)  \equiv P\left(  y;g_{2}\left(  t\right)
,g_{3}\left(  t\right)  \right)  $ which satisfy
\begin{equation}
P_{y}^{2}=4P^{3}-g_{2}P-g_{3} \label{wie6}%
\end{equation}
where $g_{2}\left(  t\right)  $, $g_{3}\left(  t\right)  $ are the invariants
of $P\left(  y,t\right)  $. In the following $C\left(  t\right)  $ is arbitrary.

\begin{center}%
\begin{tabular}
[c]{|c|c|c|c|}\hline\hline
\multicolumn{1}{||c|}{} & \multicolumn{1}{||c|}{%
\begin{tabular}
[c]{c}%
\textit{Parameters}\\
$c_{i}$%
\end{tabular}
} & \multicolumn{1}{||c|}{$\tilde{Z}\left(  \omega,t\right)  =g\left(
x\right)  e^{-\mu}$} & \multicolumn{1}{||c||}{%
\begin{tabular}
[c]{c}%
\textit{Weierstrass}\\
\textit{P-function}\\
\textit{properties}%
\end{tabular}
}\\\hline\hline
1 & $c_{0}=0$ & $-\frac{6}{b}P\left(  \omega,t\right)  +\frac{\beta}{2b}$ &
\begin{tabular}
[c]{c}%
$g_{2}=\frac{\beta^{2}}{12}$\\
$g_{3}=C\left(  t\right)  $%
\end{tabular}
\\\hline
2 &
\begin{tabular}
[c]{c}%
$c_{0}\neq0$\\
$\beta=\frac{49c_{0}^{2}}{25}$%
\end{tabular}
& $-\frac{24c_{0}^{2}}{25b}q^{2}P\left(  q,t\right)  +\frac{24c_{0}^{2}}{25b}$
&
\begin{tabular}
[c]{c}%
$q\left(  \omega\right)  =e^{\frac{2c_{0}\omega}{5}}$\\
$g_{2}=0$\\
$g_{3}=C\left(  t\right)  \neq0$%
\end{tabular}
\\\hline
3 &
\begin{tabular}
[c]{c}%
$c_{0}\neq0$\\
$\beta=\frac{49c_{0}^{2}}{25}$%
\end{tabular}
& $\frac{96c_{0}^{2}}{25b}\frac{\left(  q+1\right)  }{\left(  q+2\right)
^{2}}$ &
\begin{tabular}
[c]{c}%
$q\left(  \omega\right)  =e^{\frac{2c_{0}\omega}{5}}$\\
$g_{2}=0$, $g_{3}=0$%
\end{tabular}
\\\hline
4 &
\begin{tabular}
[c]{c}%
$c_{0}\neq0$\\
$\beta=\frac{c_{0}^{2}}{25}$%
\end{tabular}
& $-\frac{24c_{0}^{2}}{25b}q^{2}P\left(  q,t\right)  $ &
\begin{tabular}
[c]{c}%
$q\left(  \omega\right)  =e^{\frac{2c_{0}\omega}{5}}$\\
$g_{2}=0$\\
$g_{3}=C\left(  t\right)  \neq0$%
\end{tabular}
\\\hline
5 &
\begin{tabular}
[c]{c}%
$c_{0}\neq0$\\
$\beta=\frac{c_{0}^{2}}{25}$%
\end{tabular}
& $-\frac{24c_{0}^{2}}{25b}\frac{q^{2}}{\left(  q+1\right)  ^{2}}$ &
\begin{tabular}
[c]{c}%
$q\left(  \omega\right)  =e^{\frac{2c_{0}\omega}{5}}$\\
$g_{2}=0$,$g_{3}=0$%
\end{tabular}
\\\hline
\end{tabular}
\medskip

\textit{Table 2: Solutions for }$\tilde{Z}\left(  \omega,t\right)  $\textit{
when }$b\neq0$\textit{ }

\textit{with }$\beta=2c_{1}c_{3}+c_{2}^{2}$ , $x=r^{2}$
\end{center}

Using (\ref{vc51}) table 2 (entry 1) gives rise to the metric%

\begin{equation}
d\sigma^{2}=\frac{B^{2}\left(  t\right)  \left(  P_{\omega}-\eta\tilde{P}%
_{t}\right)  ^{2}}{\left(  P-\frac{\beta}{12}\right)  ^{2}}dt^{2}-\frac
{h^{2}\left(  x\right)  b^{2}}{36\left(  P-\frac{\beta}{12}\right)  ^{2}%
}\left(  dr^{2}+r^{2}d\Omega^{2}\right)  \label{metricP7}%
\end{equation}
This is an explicit form of the solution also given by Srivastava \cite{sr}
and presented implicitly by Stephani \textit{et. al.} \cite{stepbook} .When
$C\left(  t\right)  $ is contant the solutions of table 2 are those derived by
McVittie \cite{mc}. If $g_{3}=-\beta^{3}/216$ then table 2 (entry 1) may be
integrated to give the solutions of McVittie \cite{MCV33}. Solutions for the
case $g_{3}=\beta^{3}/216$ were discussed by Chakravarty \textit{et. al.}
\cite{ch}.

The metric (\ref{metricP7}) may also be considered using the Jacobi elliptic
function $\operatorname{sn}\left(  u\right)  =\operatorname{sn}\left(
u,k\right)  $ with elliptic modulus $k$, $0<k^{2}<1$. It may be shown that%
\begin{equation}
\tilde{Z}\left(  \omega\left(  x,t\right)  ,t\right)  =\frac{2h_{\pm}}%
{b}\left\{  -\frac{3}{\operatorname{sn}^{2}\left(  u,k\right)  }+1+k^{2}%
\pm\sqrt{k^{4}-k^{2}+1}\right\}  \label{sn14}%
\end{equation}
where $h=h\left(  t\right)  $, the elliptic modulus $k=k\left(  t\right)  $
and $u$ is the independent variable:%
\begin{equation}
u=\sqrt{h_{\pm}}\omega\qquad h_{\pm}=\pm\frac{\beta}{4\sqrt{k^{4}-k^{2}+1}%
}\qquad C\left(  t\right)  =\frac{4h_{\pm}^{3}\left(  k^{2}-2\right)  \left(
k^{2}+1\right)  \left(  2k^{2}-1\right)  }{27} \label{sn7}%
\end{equation}

Solutions (\ref{sn14}) with constant $C\left(  t\right)  $ have been discussed
by Sussman \cite{sus}. In particular \ when $\beta>0$ , $k=1$ then sn$\left(
u,1\right)  =\tanh\left(  u\right)  $ and solution of \cite{MCV33} is
recovered whilst for $k=0$ then sn$\left(  u,0\right)  =\sin\left(  u\right)
$ giving the solution \cite{ch}.

Table 2 (entries 2, 4) apply for all values of $c_{0}\neq0$. Thus when
$c_{3}=0$ then $\beta=c_{2}^{2}$ and entry 2 gives $c_{0}=\pm5c_{2}/7$
discussed by Srivastava \cite{sr}. When $c_{1}=0$ then again $c_{0}=\pm
5c_{2}/7$ discussed by Wyman \cite{wy}. table 2 (entries 2, 4) were first
obtained by Wyman \cite{wy} directly from (\ref{kus51}) and McVittie \cite{mc}
from equation (\ref{barb30}). Both applied techniques originally due to Ince
\cite{in}. Finally while entries 2, 4 contain two arbitrary functions of time
$\eta\left(  t\right)  $ and $C\left(  t\right)  $ table 2 (entries 3, 5)
contain only $\eta\left(  t\right)  .$ Moreover the entries 2 to 5 are known
only for particular values of $\beta$ and so further investigation of the case
$c_{0}\neq0$ is necessary.

\section{Non-comoving (non-zero shear) solutions}

\subsection{ Lie symmetry reduction, general case}

Consider the symmetries of equation (\ref{id35}) using the condition for
invariance (\ref{inv1}). In this case and without loss of generality the
solutions of the determining equations may be written in the form:%
\begin{equation}
\eta_{1}=r\qquad\eta_{2}=\eta\left(  t\right)  \qquad\qquad\phi_{1}=c_{0}%
-\eta_{t}\qquad\phi_{2}=c_{0}-1 \label{sym5}%
\end{equation}
The surface conditions (\ref{surf2}) \ may now be integrated and it is found
that%
\begin{align}
\lambda &  =\Psi\left(  \omega\right)  +c_{0}\int\frac{dt}{\eta}-\ln\eta
\qquad\qquad\mu=\Phi\left(  \omega\right)  +\left(  c_{0}-1\right)  \ln
r\nonumber\\
\omega &  =\ln r-\int\frac{dt}{\eta} \label{sym6a}%
\end{align}
Hence equation (\ref{id35}) reduces to the ordinary differential equation
\begin{equation}
e^{2\left(  \Phi-\Psi+c_{0}\omega\right)  }\left\{  \left(  2c_{0}\Psi
_{\omega}+K_{5}\right)  ^{2}+K_{6}\left(  K_{5}+2c_{0}\Psi_{\omega}-2c_{0}%
\Phi_{\omega}\right)  \right\}  -K_{5}K_{6}=0 \label{oso54}%
\end{equation}
where:%
\begin{align}
K_{5}  &  =\Phi_{\omega\omega}-\Psi_{\omega\omega}+\Phi_{\omega}^{2}%
-\Psi_{\omega}^{2}+2c_{0}\Phi_{\omega}+c_{0}^{2}-1\nonumber\\
K_{6}  &  =\Phi_{\omega\omega}+\Psi_{\omega\omega}-\Phi_{\omega}^{2}%
+\Psi_{\omega}^{2}-2\Phi_{\omega}\Psi_{\omega}-2c_{0}\Phi_{\omega}-2c_{0}%
\Psi_{\omega}-c_{0}^{2}+1=K_{1}r^{2}\nonumber\\
r\eta\left(  t\right)  K_{3}  &  =\Phi_{\omega}\Psi_{\omega}-\Phi
_{\omega\omega} \label{oso58b}%
\end{align}
Equation (\ref{oso54}) was found orginally using ad hoc mathematical
simplifications by McVittie and Wiltshire \cite{wt}. In addition using
(\ref{is40}), the pressure, energy density and mass function satisfy%
\begin{align}
8\pi p\left(  r,t\right)   &  =r^{-2c_{0}}\left\{  e^{-2\Phi}\left[
\Psi_{\omega\omega}+\Psi_{\omega}^{2}+\Phi_{\omega\omega}\right]  \right.
\nonumber\\
&  \left.  +e^{-2\left(  \Psi-c_{0}\omega\right)  }\left[  2\Phi_{\omega}%
\Psi_{\omega}-2\Phi_{\omega\omega}-3\Phi_{\omega}^{2}-2c_{0}\Phi_{\omega
}\right]  \right\}  \label{pres6}%
\end{align}%
\begin{align}
8\pi\rho\left(  r,t\right)   &  =r^{-2c_{0}}\left\{  e^{-2\Phi}\left[
\Psi_{\omega\omega}+\Psi_{\omega}^{2}-2\Phi_{\omega}\Psi_{\omega}-2c_{0}%
\Psi_{\omega}\right.  \right. \nonumber\\
&  \left.  \left.  -\Phi_{\omega\omega}-2\Phi_{\omega}^{2}-4c_{0}\Phi_{\omega
}-2c_{0}^{2}+2\right]  +3e^{-2\left(  \Psi-c_{0}\omega\right)  }\Phi_{\omega
}^{2}\right\}  \label{dens6}%
\end{align}%
\begin{equation}
2m\left(  r,t\right)  =r^{c_{0}}e^{\Phi}\left\{  e^{2\left(  \Phi+c_{0}%
\omega-\Psi\right)  }\Phi_{\omega}^{2}+1-\left(  \Phi_{\omega}+c_{0}\right)
^{2}\right\}  \label{mass6}%
\end{equation}
Also with the additional substitution%

\begin{equation}
S\left(  \omega\right)  =\frac{K_{1}\,r\,e^{\Psi}}{K_{3}\,\eta\left(
t\right)  }\qquad\qquad H\left(  \omega\right)  =\sqrt{4\,e^{2\,\Phi
+2\,c_{0}\,\omega}-S^{2}} \label{mkk5}%
\end{equation}
and from (\ref{xz50}) the non-zero components of the velocity four vector are%

\begin{equation}
u^{1}=-\frac{e^{-{\Phi}}\,S}{r^{c_{0}-1}\,H}\qquad\qquad u^{4}=\frac
{2\,\eta\left(  t\right)  \,e^{-\Psi+\Phi+2\,c_{0}\,\omega}}{r^{c_{0}}\,H}
\label{mkk10}%
\end{equation}
Moreover the expansion is, $\mathring{u}_{1}$ component of acceleration and
$\sigma_{11}$ component of the shear tensor are respectively:%
\begin{align}
\Theta &  =-\frac{e^{-{\Phi}}\,\left(  H\Psi_{\omega}\,+2\,H\,\Phi_{\omega
}-H_{\omega}+2\,c_{0}\,H\right)  \,S}{r^{c_{0}}\,H^{2}}\nonumber\\
&  -\frac{4\,e^{\Phi+2\,c_{0}\,\omega}\,\left(  \Phi_{\omega}+c_{0}\right)
}{r^{c_{0}}\,H\,S}+\frac{H_{\omega}e^{-{\Phi}}}{r^{c_{0}}\,S} \label{mkk15}%
\end{align}%
\begin{align}
\mathring{u}_{1}  &  =-\frac{8\,\left(  2\,H\,\Phi_{\omega}-H_{\omega}%
+c_{0}\,H\right)  \,e^{-\Psi+4\Phi+4\,c_{0}\,\omega}}{r\,H^{3}\,S}%
+\frac{2\,\Phi_{\omega}\,e^{-\Psi+2\,\Phi+2\,c_{0}\,\omega}}{r\,S}\nonumber\\
&  -\frac{4\,e^{2\,\Phi+2\,c_{0}\,\omega}\,\left(  H\,\Psi_{\omega}%
+H\,\Phi_{\omega}-H_{\omega}+c_{0}\,H\right)  }{rH^{3}} \label{mkk20}%
\end{align}%
\begin{align}
\sigma_{11}  &  =\frac{32\,r^{c_{0}-2}\,e^{5\,\Phi+4\,c_{0}\,\omega}\,\left(
H\,\Psi_{\omega}-H_{\omega}\right)  }{3\,H^{4}\,S}-\frac{8\,r^{c_{0}%
-2}\,e^{3\,\Phi+2\,c_{0}\,\omega}\,\left(  \Psi_{\omega}-\Phi_{\omega}%
-c_{0}\right)  }{3\,H\,S}\nonumber\\
&  +\frac{16\,r^{c_{0}-2}\,\left(  H\Phi_{\omega}-H_{\omega}+c_{0}\,H\right)
\,e^{-\Psi+5\,\Phi+4\,c_{0}\,\omega}}{3\,H^{4}} \label{mkk30}%
\end{align}

\subsection{ Some similarity solutions}

Solutions of Einstein's equation that exhibit a simple baratropic equation of
state $p=\left(  \gamma-1\right)  \rho$ are of considerable interest and from
(\ref{pres6}) and (\ref{dens6}) the resulting equation would be an ordinary
differential equation in the variable $\omega$. In principle one might attempt
to solve this simultaneously with (\ref{oso54}). However it would not
simultaneously be possible to fit the solution to empty spacetime, other than
for dust. This follows because from (\ref{pres6}) the condition \cite{bon,wlt}
for zero pressure at the boundary surface, $r_{b}=r_{b}\left(  t\right)  $
implies that $\omega=\omega_{b}$ where $\omega_{b}$ is constant. However the
boundary the mass function (\ref{mass6}) should be constant . This is not
possible except when $c_{0}=0$. Results corresponding to $c_{0}=0$ and a
particular solution of (\ref{oso54}\textit{)} for $c_{0}\neq0$ are given in
Table 3.

\begin{center}%
\begin{tabular}
[c]{|c|c|c|c|}\hline\hline
& \multicolumn{1}{||c|}{%
\begin{tabular}
[c]{c}%
\textit{Solutions of (\ref{oso54})}\\
\textit{for }$\Phi$\textit{ and }$\Psi$%
\end{tabular}
} & \multicolumn{1}{||c|}{%
\begin{tabular}
[c]{c}%
\textit{Ancillary}\\
\textit{functions }$Q\left(  \omega\right)  $\\
\textit{\& conditions}%
\end{tabular}
} & \multicolumn{1}{||c||}{%
\begin{tabular}
[c]{c}%
\textit{Constant}\\
\textit{parameters for}\\
(\ref{oso54})
\end{tabular}
}\\\hline\hline
\textit{1} &
\begin{tabular}
[c]{c}%
$\Phi=\Psi-c_{0}\omega+\ln k$\\
$\Psi=\dfrac{\ln Q\left(  \omega\right)  }{2\alpha\left(  k^{2}-1\right)  }$%
\end{tabular}
&
\begin{tabular}
[c]{l}%
$Q=(A\sinh\sqrt{\alpha}\omega+$\\
$B\cosh\sqrt{\alpha}\omega)e^{2k^{2}\alpha c_{0}\omega}$%
\end{tabular}
&
\begin{tabular}
[c]{c}%
$c_{0}\neq0$, $k\neq1,$ $\alpha>0$\\
$2\alpha=\frac{1}{1+k^{2}\left(  2c_{0}^{2}-1\right)  }$\\
\textit{Section \ref{sec1}}%
\end{tabular}
\\\hline
\textit{2} &
\begin{tabular}
[c]{c}%
$\Phi=\Psi-c_{0}\omega+\ln k$\\
$\Psi=\dfrac{\ln Q\left(  \omega\right)  }{2\alpha\left(  k^{2}-1\right)  }$%
\end{tabular}
&
\begin{tabular}
[c]{l}%
$Q=(A\sin\sqrt{-\alpha}\omega+$\\
$B\cos\sqrt{-\alpha}\omega)e^{2k^{2}\alpha c_{0}\omega}$%
\end{tabular}
&
\begin{tabular}
[c]{c}%
$c_{0}\neq0$, $k\neq1,$ $\alpha<0$\\
$2\alpha=\frac{1}{1+k^{2}\left(  2c_{0}^{2}-1\right)  }$\\
\textit{Section \ref{sec1}}%
\end{tabular}
\\\hline
\textit{3} &
\begin{tabular}
[c]{c}%
$\Phi=\Psi-c_{0}\omega$\\
$\Psi=Ae^{\frac{\omega}{c_{0}}}+Q\left(  \omega\right)  $%
\end{tabular}
&
\begin{tabular}
[c]{l}%
$Q=$\\
$\frac{3c_{0}}{4}\omega+a_{0}$%
\end{tabular}
&
\begin{tabular}
[c]{c}%
$c_{0}\neq0$, $k=1$\\
\textit{Section \ref{sec1}}%
\end{tabular}
\\\hline
\textit{4} &
\begin{tabular}
[c]{c}%
\textit{All }$\Phi$\textit{ and }$\Psi$\\
$K_{5}=0$%
\end{tabular}
&
\begin{tabular}
[c]{c}%
\textit{T-solutions}\\
$e^{2\left(  \Phi-\Psi\right)  }-1>0$%
\end{tabular}
&
\begin{tabular}
[c]{c}%
$c_{0}=0$\\
\textit{Section \ref{sec2}}%
\end{tabular}
\\\hline
\textit{5} &
\begin{tabular}
[c]{c}%
\textit{All }$\Phi$\textit{ and }$\Psi$\\
$\dfrac{K_{5}}{K_{6}}=e^{2\left(  \Psi-\Phi\right)  }-1$%
\end{tabular}
&
\begin{tabular}
[c]{c}%
\textit{R-solutions}\\
$e^{2\left(  \Phi-\Psi\right)  }-1<0$%
\end{tabular}
&
\begin{tabular}
[c]{c}%
$c_{0}=0$\\
\textit{Section \ref{sec3}}%
\end{tabular}
\\\hline
\end{tabular}
\medskip

\textit{Table 3: The functions }$\Phi$\textit{ and }$\Psi$\textit{ defining
non-comoving,}

\textit{ Lie solutions (\ref{sym6a})}
\end{center}

\subsubsection{\label{sec1}Kinematics for table 3 (entries 1,2,3)}

The non-zero components of the velocity four vector are:%

\begin{equation}
u^{1}=-\frac{e^{-{\Psi+c\omega}}\,S}{k\xi^{c_{0}-1}\,H}\qquad\qquad
u^{4}=\frac{2k\,\eta\left(  t\right)  \,e^{-\Psi+\,c_{0}\,\omega}}{\xi^{c_{0}%
}\,H} \label{bg5}%
\end{equation}
where%
\begin{equation}
S=-\frac{2\,k^{2}\,e^{\Psi}\,\left(  2\,c_{0}\,\Psi_{\omega}-1\right)
}{2\,c_{0}\,k^{2}\,\Psi_{\omega}-2\,c_{0}^{2}\,k^{2}-1}\qquad\qquad
H=\sqrt{4\,k^{2}\,e^{2\,\Psi}-S^{2}} \label{bg10}%
\end{equation}
In this way it may be shown that the kinematic parameters are:%
\begin{equation}
\Theta=-\frac{e^{c_{0}\,\omega-\Psi}\,\left(  4\,k^{2}\,e^{2\,\Psi}%
\,S_{\omega}-3\,\Psi_{\omega}\,S^{3}+8\,k^{2}\,e^{2\,\Psi}\,\Psi_{\omega
}S\right)  }{k\,\xi^{c_{0}}\,H^{3}} \label{bg15}%
\end{equation}%
\begin{align}
\xi\,H^{4}\mathring{u}_{1}  &  =4k^{2}\,e^{2\,\Psi}\,\left(  2\Psi_{\omega
}\,S^{2}-1\,S\,S_{\omega}\right)  +8k^{4}\,e^{3\,\Psi}\,\left(  \,c_{0}%
\,S-S_{\omega}\right) \label{bg20}\\
&  +2k^{2}\,e^{\Psi}\,\left(  \,\Psi_{\omega}-\,c_{0}\right)  \,S^{3}%
-16\,k^{4}\,e^{4\,\Psi}\Psi_{\omega}\nonumber
\end{align}

\begin{equation}
\sigma_{1,1}=\frac{16\,k^{5}\,\xi^{c_{0}-4}\,e^{4\,\Psi-c_{0}\,\omega
}\,\left(  S+2\,e^{\Psi}\right)  \,\left(  S_{\omega}-\Psi_{\omega}\,S\right)
}{9\,H^{10}\,S} \label{bg25}%
\end{equation}
where:%

\begin{equation}
\Psi_{\omega\omega}=2\alpha\,\left(  1-k^{2}\,\right)  \Psi_{\omega}%
^{2}+4\,c_{0}\,\alpha\,k^{2}\,\Psi_{\omega}-\,\alpha\,k^{2}-\frac{1}{2}
\label{bg30}%
\end{equation}

\begin{equation}
S_{\omega}=-\frac{2\,k^{2}\,e^{\Psi}\,\left(  4\,c_{0}^{2}\,k^{2}%
\,\Psi_{\omega}^{3}-4c_{0}\,\left(  c_{0}^{2}\,k^{2}+1\right)  \,\Psi_{\omega
}^{2}+\left(  1-2\,c_{0}^{2}\,k^{2}\right)  \,\Psi_{\omega}+2\,c_{0}%
^{3}\,k^{2}+c_{0}\right)  }{\left(  2\,c_{0}\,k^{2}\Psi_{\omega}-2\,c_{0}%
^{2}\,k^{2}-1\right)  ^{2}} \label{bg35}%
\end{equation}

Note that solutions of table 3 (entries 1,3) have been presented by McVittie
and Wiltshire \cite{wt} with physical properties described by Knutsen
\cite{kn1,kn}. The case of table 3 (entry 3) with $A=0$ gives rise to a stiff
equation of state $p=\rho$ and the metric may be transformed to a comoving
system resulting in the static solution of Tolman \cite{to}. Table 3 (entry 2)
does not appear to have been been presented in the literature.

\subsubsection{\label{sec2}Discussion of table 3 (entry 4), T-solutions,
kinematics}

Consider (\ref{oso54}) with $c_{0}=0$ with $K_{5}=0$. Hence the metric is:%
\begin{equation}
d\sigma^{2}=e^{2\Psi}\eta^{-2}dt^{2}-e^{2\Phi}r^{-2}\left(  dr^{2}%
+r^{2}d\Omega^{2}\right)  \label{met5}%
\end{equation}
From (\ref{mass6}) to (\ref{mkk10}) and applying $K_{5}=0$ it follows that:%
\begin{equation}
\frac{\Phi_{\omega}^{2}}{\dfrac{2m\left(  r,t\right)  }{e^{\Phi}}-1}%
=\frac{\left(  u^{1}\right)  ^{2}}{r^{2}e^{-2\Phi}}=\frac{\left(
u^{4}\right)  ^{2}}{\eta^{2}e^{-2\Phi-4\Psi}}=\frac{1}{e^{2\left(  \Phi
-\Psi\right)  }-1}>0 \label{uva}%
\end{equation}
The expressions for (\ref{pres6}) and (\ref{dens6}) now depend only upon
$\omega$. On defining the timelike variable $\tau$ and spacelike variable
according to
\begin{equation}
\tau=e^{\Phi}\qquad\qquad q=\ln r+\int\frac{d\omega}{e^{2\left(  \Phi
-\Psi\right)  }-1} \label{time}%
\end{equation}
then (\ref{met5}) may be transformed orthogonally to the form:%
\begin{align}
d\sigma^{2}  &  =e^{2V}d\tau^{2}-e^{2W}dq^{2}-\tau^{2}d\Omega^{2}\nonumber\\
e^{2V}  &  =\frac{1}{\Phi_{\omega}^{2}\left(  e^{2\left(  \Phi-\Psi\right)
}-1\right)  }\qquad\qquad e^{2W}=e^{2\Psi}\left(  e^{2\left(  \Phi
-\Psi\right)  }-1\right)  \label{guva}%
\end{align}
and so $V$ and $W$ depend on $\tau$ alone. The isotropy condition $\tilde
{G}_{1}^{1}=\tilde{G}_{2}^{2}$ for the metric (\ref{guva}):%
\begin{equation}
W_{\tau\tau}+W_{\tau}^{2}-V_{\tau}W_{\tau}-\frac{1}{\tau^{2}}+\frac{\left(
W_{\tau}+V_{\tau}\right)  }{\tau}-\frac{e^{2V}}{\tau^{2}}=0 \label{iso26}%
\end{equation}

The region of spaceime defined by (\ref{guva}), (\ref{iso26}) is the T-region,
Novikov \cite{nv1}. Some solutions of (\ref{iso26}) have been presented by
McVittie and Wiltshire \cite{rjw1}.

It may be shown using (\ref{xz12}) that the expansion, acceleration and
non-zero components of the shear tensor for (\ref{guva}) are given by:%
\begin{equation}%
\begin{tabular}
[c]{ll}%
$\Theta=\frac{\left(  \tau W_{\tau}+2\right)  e^{-V}}{\tau}$ & $\qquad
\mathring{u}_{i}=0\qquad\forall i$\\
$\sigma_{11}=-\dfrac{2\left(  \tau W_{\tau}-1\right)  e^{2W-V}}{3\tau}$ &
$\qquad\sigma_{22}=\dfrac{\sigma_{33}}{\sin^{2}\theta}=\dfrac{\tau\left(  \tau
W_{\tau}-1\right)  e^{-V}}{3}$%
\end{tabular}
\ \label{tsh}%
\end{equation}
giving the shear invariant:%

\begin{equation}
\sigma=\dfrac{2\left(  \tau W_{\tau}-1\right)  ^{2}e^{-2V}}{3\tau^{2}}
\label{sheari}%
\end{equation}

Note that the vacuum solution of Einstein's equations corresponding to the
metric (\ref{met5}) is given in terms of the Weierstrass $P-$function
$P\left(  \omega\right)  $ satisfying the equation $P_{\omega}^{2}%
=4P^{3}-\frac{P}{12}+36M^{2}+\frac{1}{216}$ where $2M$ is the Schwarzchild
radius. Using (\ref{guva}) the vacuum solution is thus:%
\begin{equation}
\frac{e^{2\Psi}}{P_{\omega}^{2}}=\frac{e^{2\Phi}}{36M^{2}}=\frac{1}{\left(
1-12P\right)  ^{2}}\qquad\Rightarrow\qquad e^{2W}=e^{-2V}=\frac{2M}{\tau}-1
\label{soln5}%
\end{equation}

\subsubsection{\label{sec3}Discussion of table 3 (entry 5)$,$ R-solutions}

Now consider (\ref{oso54}) with $c_{0}=0$. Hence $K_{6}-e^{2\left(  \Phi
-\Psi\right)  }\left(  K_{5}+K_{6}\right)  =0$ so that by (\ref{oso58b})
$rK_{1}=-2K_{3}\eta e^{2\left(  \Phi-\Psi\right)  }$. The metric is again
(\ref{met5}) but now (\ref{mass6}), (\ref{mkk10}) give:%
\begin{equation}
\frac{e^{-2\Psi}\Phi_{\omega}^{2}}{1-\dfrac{2m\left(  r,t\right)  }{e^{\Phi}}%
}=\frac{\left(  u^{1}\right)  ^{2}}{r^{2}}=\frac{\left(  u^{4}\right)  ^{2}%
}{\eta^{2}}=\frac{1}{e^{2\Psi}-e^{2\Phi}}>0 \label{zk4a}%
\end{equation}
On defining the spacelike variable, $q$ and a timelike variable according to%
\begin{equation}
q=e^{\Phi}\qquad\text{\textit{and}}\qquad\tau=\ln r+\int\frac{d\omega
}{1-e^{2\left(  \Phi-\Psi\right)  }} \label{sp40}%
\end{equation}
then the orthogonal metric is again (\ref{met5}) but now%
\begin{equation}
e^{2V}=e^{2\Psi}\left(  1-e^{2\left(  \Phi-\Psi\right)  }\right)  \qquad\qquad
e^{2W}=\frac{1}{\Phi_{\omega}^{2}\left(  1-e^{2\left(  \Phi-\Psi\right)
}\right)  } \label{sp51}%
\end{equation}
and so $V$ and $W$ depend on $q$ alone. The isotropy condition $\tilde{G}%
_{1}^{1}=\tilde{G}_{2}^{2}$ corresponding to the metric (\ref{met5}) is now
the familiar%
\begin{equation}
V_{qq}+V_{q}^{2}-V_{q}W_{q}-\frac{1}{q^{2}}-\frac{\left(  W_{q}+V_{q}\right)
}{q}+\frac{e^{2W}}{q^{2}}=0 \label{ex16}%
\end{equation}
The region of space-time defined by (\ref{sp51}), (\ref{ex16}) is the
R-region, Novikov \cite{nv1}.

\section{Non-comoving, non-Lie cases with shear}

\subsection{Introduction and symmetry reduction}

The non-classical method is a generalisation of the classical Lie group
approach due to Bluman and Cole \cite{blcl} that incorporates the surface
invariant condition (\ref{surf2}) into the condition of form invariance
(\ref{inv1}). Thus non-classical symmetries may be found by solving the
non-linear set of determining equations%

\begin{equation}
\mathcal{X}_{E}^{(2)}\left(  \boldsymbol{\Delta}\right)  \mathbf{\mid
}_{\boldsymbol{\Delta=0,\Omega=0}}=0 \label{nnc}%
\end{equation}
where $\boldsymbol{\Omega=}\left\{  \Omega_{1}\text{,}\Omega_{2}\right\}
$.The non-classical method is one example of a more general conditional
symmetry approach described by \ Ibragimov \cite{ib} in which the condition
(\ref{inv1}) for form invariance is supplemented by an additional condition.

However, the complexity of this method and the condition of isotropy for an
non-comoving observer (\ref{id35}) present prohibitive practical difficulties.
Instead a restricted nonclassical approach, namely the direct method of
Clarkson and Kruskal \cite{clk} will be employed. This lacks the generality of
the non-classical method \cite{nc} but is easier to implement. The approach
does not involve group theory but rather imposes a given similarity ansatz on
the differential equations in question. In the context of this problem the
ansatz (\ref{arb23}) and (\ref{arb24}) for the comoving isotropy equations
will be used. Thus:%
\begin{equation}
\lambda=\Psi\left(  \omega\right)  +\ln A\left(  t\right)  \qquad\qquad
\mu=\Phi\left(  \omega\right)  +h\left(  r\right)  \label{rep5}%
\end{equation}
where%
\begin{equation}
\omega=\int\frac{dr}{\frac{c_{3}}{r}+c_{2}r-\frac{c_{1}r^{3}}{2}}-\int
\frac{dt}{\eta\left(  t\right)  }\qquad h\left(  r\right)  =\int\frac
{c_{0}-c_{2}+c_{1}r^{2}}{\frac{c_{3}}{r}+c_{2}r-\frac{c_{1}r^{3}}{2}%
}dr\label{rep10}%
\end{equation}
will be applied directly to the general equation (\ref{id35}). In this way the
isotropy condition becomes:%
\begin{align}
&  2\,K_{7}\,\eta^{2}\left(  t\right)  \,r^{2}\,\left(  2\,\Phi_{\omega}%
\Psi_{\omega}+2\,c_{0}\,\Psi_{\omega}-2\,\Phi_{\omega\omega}+K_{7}%
-8\,c_{1}\,c_{3}\right)  \nonumber\\
&  +K_{7}\,e^{-2\,\Psi+2\,\Phi-2\,A+2\,c_{0}\,\int{\frac{1}{\eta\left(
t\right)  }}{\,dt}+2\,c_{0}\,\omega}\,\left(  \Phi_{\omega}\Psi_{\omega}%
-\Phi_{\omega\omega}-\left(  \eta\,A_{t}+\eta_{t}\right)  \,\Phi_{\omega
}\right)  \nonumber\\
&  +2\,e^{-2\,\Psi+2\,\Phi-2\,A+2\,c_{0}\,\int{\frac{1}{\eta\left(  t\right)
}}{\,dt}+2\,c_{0}\,\omega}\,\left(  \Phi_{\omega}\,\Psi_{\omega}-\Phi
_{\omega\omega}\right)  ^{2}\nonumber\\
&  +4\,c_{3}\,K_{7}\,\eta^{2}\,\left(  \Psi_{\omega}-2\,\Phi_{\omega}%
+2\,c_{2}-2\,c_{0}\right)  \nonumber\\
&  +2\,c_{1}\,K_{7}\,\eta^{2}\,r^{4}\,\left(  \Psi_{\omega}-2\Phi_{\omega
}-2\,c_{2}-2\,c_{0}\right)  =0\label{isot9}%
\end{align}
where%
\begin{align}
K_{7} &  =\Psi_{\omega\omega}+\Psi_{\omega}^{2}-2\,\Phi_{\omega}\Psi_{\omega
}-2\,c_{0}\Psi_{\omega}+\Phi_{\omega\omega}\label{k777}\\
&  -\Phi_{\omega}^{2}-2\,c_{0}\Phi_{\omega}+2\,c_{1}\,c_{3}+c_{2}^{2}%
-c_{0}^{2}\nonumber
\end{align}
Inspection of these equations reveals that $\eta(t)\sim1/t$ and $\eta(t)\sim
$\textit{constant.} give rise to solutions of Einstein's equations. These
solutions are summarised in four cases presented in table 4.

\subsection{Some particular solutions of (\ref{isot9})}

In table 4 $c_{3}=\frac{1}{2}$, $A=1$ are chosen without any loss in generality.

\begin{center}%
\begin{tabular}
[c]{|c|c|c|c|c|}\hline\hline
& \multicolumn{1}{||c|}{$\boldsymbol{e}^{2\Psi}$} &
\multicolumn{1}{||c|}{$\boldsymbol{e}^{2\Phi}$} & \multicolumn{1}{||c|}{%
\begin{tabular}
[c]{l}%
\textit{Ancillary }\\
\textit{functions}%
\end{tabular}
} & \multicolumn{1}{||c||}{%
\begin{tabular}
[c]{c}%
\textit{Constant }\\
\textit{parameters}%
\end{tabular}
}\\\hline\hline
\textit{1} & $k_{1}Q^{2}-c_{10}Q$ & $Q$ &
\begin{tabular}
[c]{c}%
$Q=\dfrac{H^{\frac{2}{3}}-c_{10}+c_{10}^{2}H^{-\frac{2}{3}}}{k_{1}}$\\
$2H=\sqrt{\left(  c_{5}\omega+c_{6}\right)  ^{2}+4c_{10}^{3}}$\\
$+c_{5}\omega+c_{6}$\\
$\eta\left(  t\right)  =-\frac{1}{2c_{10}t}$\\
$\omega=r^{2}+c_{10}t^{2}$%
\end{tabular}
&
\begin{tabular}
[c]{l}%
$c_{0},c_{1},c_{2}=0,$\\
$c_{3}=\frac{1}{2},A=1$\\
\textit{Section \ref{sec4}}%
\end{tabular}
\\\hline
\textit{2} & $\boldsymbol{e}^{2\Psi}$ & $\boldsymbol{e}^{2\Psi}$ &
\begin{tabular}
[c]{c}%
$\eta(t)=\frac{1}{2t}$\\
$\omega=r^{2}-t^{2}$%
\end{tabular}
&
\begin{tabular}
[c]{l}%
$c_{0},c_{1},c_{2}=0,$\\
$c_{3}=\frac{1}{2},A=1$\\
\textit{Section \ref{sec4}}%
\end{tabular}
\\\hline
\textit{3} & $z^{4}$ & $e^{2\omega}z^{2}$ &
\begin{tabular}
[c]{l}%
$z=\left(  c_{5}+c_{6}e^{2\omega}\right)  ^{\frac{1}{3}}$\\
$\eta\left(  t\right)  =-\frac{1}{b}$\\
$e^{\omega}=\sqrt{r^{2}+c_{3}}e^{bt}$%
\end{tabular}
&
\begin{tabular}
[c]{l}%
$c_{0},c_{1}=0$\\
$c_{2},A=1,$\\
\textit{Section \ref{sec5}}%
\end{tabular}
\textit{ }\\\hline
\textit{4} & $z^{\frac{4}{3}}$ & $z^{\frac{2}{3}}$ &
\begin{tabular}
[c]{l}%
$z=A+b\omega$\\
$\eta\left(  t\right)  =-\frac{1}{b}$\\
$\omega=r^{2}+bt$%
\end{tabular}
- &
\begin{tabular}
[c]{l}%
$c_{0},c_{1},c_{2}=0,$\\
$c_{3}=\frac{1}{2},A=1$\\
\textit{Section \ref{sec5}}%
\end{tabular}
\textit{ }\\\hline
\end{tabular}
\medskip

\textit{Table 4: The functions }$\Phi$\textit{ and }$\Psi$ \textit{defining}
\textit{non-comoving,}

\textit{ non-Lie solutions of (\ref{isot9})}
\end{center}

\subsubsection{ \label{sec4}Kinematics for table 4 (entries 1 and 2)}

In the case of table 4 (entry 1) it may be shown that
\begin{equation}
Q_{\omega}=\frac{2\,c_{5}\,\left(  H^{\frac{4}{3}}-c_{10}^{2}\right)
\,H^{\frac{1}{3}}\,}{3\,k_{1}\,\left(  H^{2}+c_{10}^{3}\right)  } \label{ww12}%
\end{equation}
The pressure, density and mass function are found to be:%
\begin{equation}
8\pi p=-\frac{Q_{\omega}\,\left(  5\,k_{1}\,r^{2}\,Q_{\omega}\,-7\,c_{10}%
\,\omega Q_{\omega}-6\,k_{1}\,Q^{2}+6\,c_{10}\,Q\right)  }{Q^{3}\,\left(
k_{1}\,Q-c_{10}\right)  } \label{ww17j}%
\end{equation}%
\begin{equation}
8\pi\rho=-\frac{Q_{\omega}\,\left(  5\,k_{1}\,r^{2}\,Q\,Q_{\omega}%
-3\,c_{10}\,\omega\,Q_{\omega}+6\,k_{1}\,Q^{2}-6\,c_{10}\,Q\right)  }%
{Q^{3}\,\left(  k_{1}\,Q-c_{10}\right)  } \label{ww19}%
\end{equation}%
\begin{equation}
m=-\frac{r^{3}\,Q_{\omega}\left(  k_{1}\,r^{2}\,QQ_{\omega}-c_{10}%
\,\omega\,Q_{\omega}+2\,k_{1}\,Q^{2}-2\,c_{10}\,Q\right)  }{2\,Q^{\frac{3}{2}%
}\,\left(  k_{1}\,Q-c_{10}\right)  } \label{ww21}%
\end{equation}
and from (\ref{xz50}) the non-zero components of the velocity four vector
\begin{equation}
u^{1}=\frac{c_{10}\,\eta\left(  t\right)  u^{4}}{r}\qquad\qquad u^{4}=\frac
{r}{\sqrt{Q\left(  k_{1}\,r^{2}\,Q-c_{10}\,\omega\right)  }} \label{ww23a}%
\end{equation}
The expansion is%

\begin{align}
\Theta &  =\frac{2\,c_{10}\,t\,r\,\left(  3\,k_{1}^{2}\,r^{2}\,Q^{2}%
-2\,k_{1}\,c_{10}\,r^{2}\,Q-4\,k_{1}\,c_{10}\,\omega\,Q+3\,c_{10}^{2}%
\,\omega\right)  \,Q_{\omega}}{Q^{\frac{3}{2}}\,\left(  k_{1}\,Q-c_{10}%
\right)  \,\left(  k_{1}\,r^{2}\,Q-c_{10}\,\omega\right)  ^{\frac{3}{2}}%
}\nonumber\\
&  +\frac{c_{10}\,t\,\left(  k_{1}\,r^{2}\,Q+2\,c_{10}\,r^{2}-2\,c_{10}%
\,\omega\right)  }{r\,\sqrt{Q}\,\left(  k_{1}\,r^{2}\,Q-c_{10}\,\omega\right)
^{\frac{3}{2}}} \label{ww26a}%
\end{align}
The component $\mathring{u}_{1}$ of acceleration is
\begin{align}
\mathring{u}_{1}  &  =-\frac{c_{10}\,r\,\left(  2\,\xi^{2}-\omega\right)
\,\left(  k_{1}\,Q-c_{10}\right)  }{\left(  k_{1}\,r^{2}\,Q-c_{10}%
\,\omega\right)  ^{2}}\nonumber\\
&  -\frac{r\,\left(  2\,k_{1}^{2}\,r^{4}\,Q^{2}-3\,k_{1}\,c_{10}%
\,\omega\,r^{2}\,Q+2\,c_{10}^{2}\,\omega\,r^{2}-c_{10}^{2}\,\omega^{2}\right)
\,Q_{\omega}}{Q\,\left(  k_{1}\,r^{2}\,Q-c_{10}\,\omega\right)  ^{2}}
\label{ww31s}%
\end{align}
and the component $\sigma_{11}$ of the shear tensor is given through%
\begin{equation}
\sigma_{11}=-\frac{2\,c_{10}\,r\,\sqrt{\frac{\omega-r^{2}}{c_{10}}}\,\sqrt
{Q}\,W}{3\,\left(  k_{1}\,r^{2}\,Q-c_{10}\,\omega\right)  ^{\frac{5}{2}}}
\label{bgg7}%
\end{equation}
where%
\begin{align}
W  &  =2\,k_{1}\,c_{10}\,r^{4}\,Q_{\omega}-2\,k_{1}\,c_{10}\,\omega
\,r^{2}\,Q_{\omega}-2\,k_{1}^{2}\,r^{2}\,Q^{2}\nonumber\\
&  +4\,k_{1}\,c_{10}\,r^{2}\,Q+k_{1}\,c_{10}\,\omega\,Q-2\,c_{10}^{2}%
\,r^{2}-c_{10}^{2}\,\omega\label{ww37a}%
\end{align}

The case of table 4 (entry 2) with $c_{10}=-1$. is the conformally Minkowskian
case where $\Phi$ $=\Psi$ are arbitrary functions of $\omega$. However it has
been shown see for example Infeld and Schild \cite{if} and Tauber \cite{ta}
that the hyperbolic case $k=-1$ of the Robertson Walker metric may be
transformed into table 4 (entry 2) and so this case will not be pursued further.

\subsubsection{\textit{ \label{sec5}}Kinematics for table 4 (entries 3 and 4)}

In the case of table 4 (entry 3) with $c_{2}=1$ (without loss of generality)
it is convenient to define the following additional functions.%

\begin{align}
W  &  =\left(  6\,r^{2}\,z^{3}+3\,c_{3}\,z^{3}-4\,c_{6}\,e^{2\,\omega}%
\,r^{2}\right)  \,Y-6\,r^{4}\,z^{5}+3\,c_{3}\,r^{2}\,z^{5}+4\,c_{6}%
\,e^{2\,\omega}\,r^{4}\,z^{2}\nonumber\\
Y  &  =r^{2}\,z^{2}-b^{2}\,e^{2\,\omega}\,r^{2}-c_{3}\,b^{2}\,e^{2\,\omega}
\label{kp6t}%
\end{align}
It may be shown that the pressure, density and mass function are%

\begin{equation}
8\pi p=\frac{4\,c_{6}\,\left(  9z^{3}\,\left(  r^{2}\,+\,c_{3}\,\right)
-5\,c_{6}\,e^{2\,\omega}\,r^{2}\right)  }{9\,\left(  r^{2}+c_{3}\right)
\,z^{8}}-\frac{b^{2}\,\left(  27\,z^{6}+36\,c_{6}\,e^{2\,\omega}%
\,z^{3}-28\,c_{6}^{2}\,e^{4\,\omega}\right)  }{9\,z^{10}} \label{pr52}%
\end{equation}%
\begin{equation}
8\pi\rho=\frac{b^{2}\,\left(  3\,z^{3}+2\,c_{6}\,e^{2\,\omega}\right)  ^{2}%
}{3\,z^{10}}-\frac{4\,c_{6}\,\left(  9\,r^{2}\,z^{3}+9\,c_{3}\,z^{3}%
+5\,c_{6}\,e^{2\,\omega}\,r^{2}\right)  }{9\,\left(  r^{2}+c_{3}\right)
\,z^{8}} \label{kp160}%
\end{equation}%
\begin{equation}
m=\frac{b^{2}\,e^{3\,\omega}\,r^{3}\,\left(  3\,z^{3}+2\,c_{6}\,e^{2\,\omega
}\right)  ^{2}}{18\,\left(  r^{2}+c_{3}\right)  ^{\frac{3}{2}}\,z^{7}}%
-\frac{2\,c_{6}\,e^{3\,\omega}\,r^{3}\,\left(  3\,r^{2}\,z^{3}+3\,c_{3}%
\,z^{3}+c_{6}\,e^{2\,\omega}\,r^{2}\right)  }{9\,\left(  r^{2}+c_{3}\right)
^{\frac{5}{2}}\,z^{5}} \label{kp170}%
\end{equation}
and from (\ref{xz50}) the non zero components of the velocity four vector are%
\begin{equation}
u^{4}=\frac{r}{z\sqrt{Y}}\qquad u^{1}=\frac{b\left(  r^{2}+c_{3}\right)
}{z\sqrt{Y}} \label{kp20}%
\end{equation}
The expansion is:%

\begin{equation}
\Theta=\frac{b\,\left(  18\,r^{2}\,z^{3}\,Y+9\,c_{3}\,z^{3}\,Y+12\,c_{6}%
\,e^{2\,\omega}\,r^{2}\,Y-W\right)  }{3\,r\,z^{4}\,Y^{\frac{3}{2}}}
\label{kp25}%
\end{equation}
while the non zero components of acceleration are%
\begin{align}
-\frac{\,\left(  r^{2}+c_{3}\right)  }{r}\mathring{u}_{1}  &  =\frac{\left(
-3\,r^{2}\,z^{3}+3c_{3}\,\,z^{3}+10\,c_{6}\,e^{2\,\omega}\,r^{2}\right)  }%
{Y}\nonumber\\
&  +\frac{\left(  6\,\,r^{4}\,z^{4}-3c_{3}\,\,r^{2}\,z^{4}-4\,c_{6}%
\,\,e^{2\,\omega}\,r^{4}\,z\right)  }{Y^{2}}-\left(  \frac{2\,c_{6}%
\,\,e^{2\,\omega}}{3\,z^{3}}+1\right)  \label{kp35}%
\end{align}
and the nonzero component $\sigma_{11}$ of the stress tensor is%

\begin{equation}
\sigma_{11}=\frac{2\,b\,e^{2\,\omega}\,r\,W}{9\,\left(  r^{2}+c_{3}\right)
\,Y^{\frac{5}{2}}} \label{bgg8}%
\end{equation}
Finally the case of table 4 (entry 4) when $\mathbf{c}_{2}\mathbf{=0}$ is a
solution originally due to McVittie \& Wiltshire \cite{wt} with kinematics and
other physical properties analysed by Bonnor and Knutsen \cite{br}.

\section{Conclusion.}

It has been possible to show that many solutions of Einstein's equations for
perfect fluid spheres are obtainable by the classical symmetry method. This is
not only true for the well known spheres described by a comoving observer but
it is also true for non-comoving cases. However these produce a more
restricted range of symmetries and it proved necessary to employ non-Lie
methods to extend the range of similarity solutions. Application of the
non-classical method in its full form proved impracticable for non-comoving
cases. However a restricted implementation in the form of the direct method
enabled certain new solutions of Einstein's equations to be found. The
physical and kinematical properties of these solutions have been presented in
most cases but a detailed analysis of their physical applicability will be
considered elsewhere.

Whilst the classical symmetry method has been applied succesfully in the
derivation of solutions for a comoving four velocity the analysis is far from
complete as equation (\ref{kus12}), Abel's differential equation of the second
kind requires further research. Moreover, non-classical techniques have not
yet been applied to comoving systems and certainly more work is necessary into
non-Lie approaches for non-comoving cases. In addition solutions (whether
comoving or non-comoving) that admit a simple baratropic equation of state of
the form $p=\left(  \gamma-1\right)  \rho$ \ are in general as elusive as ever.

This paper has focussed upon symmetry reduction for line elements with
spherically symmetry and certainly the methodology could be applied to find
exact solutions for perfect fluids with for example, plane or hyperbolic
symmetries. These ideas will be discussed elsewhere.

\end{document}